# The role of the electrolysis and enzymatic hydrolysis in the enhancement of the electrochemical properties of 3D-printed carbon black/poly(lactic acid) structures

Adrian Koterwa[1,#], Iwona Kaczmarzyk[2,#], Szymon Mania[3], Mateusz Cieślik[3,4], Robert Tylingo[3], Tadeusz Ossowski[1], Robert Bogdanowicz[2], Paweł Niedziałkowski[1], Jacek Ryl[4,*]

[1] Department of Analytical Chemistry, University of Gdansk, Wita Stwosza 63, 80-308 Gdansk, Poland
[2] Department of Metrology and Optoelectronics and Advanced Materials Center, Gdansk University of Technology, Narutowicza 11/12, 80-233 Gdansk, Poland
[3] Faculty of Chemistry, Gdansk University of Technology, Narutowicza 11/12, 80-233 Gdansk, Poland
[4] Institute of Nanotechnology and Materials Engineering and Advanced Materials Center, Gdansk University of Technology, Narutowicza 11/12, 80-233 Gdansk, Poland
＊ Correspondence: Jacek Ryl – jacek.ryl@pg.edu.pl
# these authors contributed equally to the manuscript

**Abstract:** Additive manufacturing, called 3D printing, starts to play an unprecedented role in developing many applications in industrial or personalized products. The conductive composite structures require additional treatment to achieve an electroactive surface useful for electrochemical devices. In this paper, the surfaces of carbon black/poly(lactic acid) CB-PLA printouts were activated by electrolysis or enzymatic digestion with proteinase K, or a simultaneous combination of both. Proposed modification protocols allowed for tailoring electrochemically active surface area and electron transfer kinetics determined by electrochemical techniques (CV, EIS) with $[Fe(CN)_6]^{4-/3-}$ redox probe. The X-ray photon spectroscopy and SEM imaging were applied to determine the delivered surface chemistry. The CB-PLA hydrolysis in alkaline conditions and under anodic polarization greatly impacts the charge transfer kinetics. The enzymatic hydrolysis of PLA with proteinase K has led to highly efficient results yet requiring an unsatisfactory prolonged activation duration of 72 h, efficiently reduced by the electrolysis carried out in the presence of the enzyme. Our studies hint that the activation protocol originates from surface electropolymerization rather than synergistic interaction between electrolysis and enzymatic hydrolysis. The detailed mechanism of CB-PLA hydrolysis supported by electrolysis has been elaborated since it pawed a new route towards a time-efficient and environmentally-friendly activation procedure.

## 1. Introduction

The advent of Additive Manufacturing (AM), also called three-dimensional printings, which is a layer by layer manufacturing fabrication process, has an unprecedented impact on development in various fields, including science, engineering, and industry. Nonetheless, affordable desktop three-dimensional printers, which have used fused deposition modeling (FDM), have become a popular tool for fabricating personalized consumer products. Fused deposition modeling or fused filament fabrication (FFF) is one of the most commonly utilized and commercially successful method by consequence of their low cost, easy-to-use interface and graphic software. Furthermore, the expiration of the original FDM patents led to growing extrusion-based printers. The development of desktop extrusion-based printers does open up new ways to approach areas including biomedical sensors [1], electronics, chemistry [2], pharmaceutics [3], medicine [4], due to least expensive to own and operate. Furthermore, they are eco-friendly, fast prototyping, and giving the possibility to design objects, which have a specific shape.

Polylactic acid (PLA), is one of the most promising materials used nowadays. It is biodegradable, recyclable, highly processable, and degrades into non-toxic products [5,6]. High conductive fillers such as graphene, carbon black (CB), and carbon nanotubes (CNT), dispersed in the thermoplastic polymer matrix, are used to achieve electrically conductive filaments. Development in FFF and commonly available conductive filaments can fabricate sensors for various applications, flow cells [7], and microfluidic devices [8]. 3D printed electrodes have been used for numerous electrochemical applications, including sensing dopamine, catechol, and 2,5,6-trinitrotoluene (TNT) [9], glucose, and simultaneous determination of uric acid and nitrile [10], energy storage devices [11], etc. Among the various materials, one of the most popular is carbon black/PLA (CB-PLA) conductive composite. Carbon Black is a material made of finely divided carbons produced by the thermal decomposition of a hydrocarbon. Conductive carbon particles in CBs are chemically bound and form agglomerates by weak Van der Waals interactions [12]. The conductivity of CB-PLA is dominated by percolation, which means that the addition of conductive fillers has little effect until enough filler is present to form a continuous particle pathway through the material. Recently research presented that 3D-printed CB/PLA electrodes have promising applications of electrochemical sensors, including fuel electroanalysis [13], electrochemical cell for detection mercury ions [14], caffeine and glucose [15], simultaneous determination of cadmium and lead ions in biological fluids [16], etc.

New carbon/polymer 3D-printout electrodes present poor electrochemical properties until surface activation treatment is performed. The activation treatments remove the excess of the polymeric matrix and expose conductive fillers at the electrode surface. It has been well documented that the chemical activation of carbon/polymer electrodes improves the efficiency of electrochemical processes. Palenzuela et al. reported the surface treatment of graphene/PLA by immersion in dimethylformamide (DMF) for 10 min to expose the graphene fillers [31] successfully. However, the most commonly used DMF has a negative impact on the environment. Therefore, other activation surface treatments are being investigated. One of them is electrochemical treatment in NaOH, which has highly improved the electrochemical response significantly. For example, Rocha et al. [16] studied the electron transfer after surface treatment in 0.5 mol L$^{-1}$ NaOH by applying a constant potential of 1.4 V for 200s and then -1.0 V for 200s. The proposed electrochemical surface treatment by Rocha et al. was used in an entirely additively manufactured electrochemical sensing platform [17]. Furthermore, Browne et al. demonstrated a combined DMF treatment and electrochemical activation process improves the activity of the 3D-printed graphene/PLA electrodes. Solvent treatments were performed by soaking the 3D-printed electrodes in DMF for 10 min, which was completed before an electrochemical activation. Then, electrochemical activation of the 3D electrodes was carried out in phosphate buffer solution (pH 7.2) using a chronoamperometry method at high oxidizing potentials over a range of times [18]. PLA, belonging to aliphatic polyesters, is usually hydrolyzed by esterases, lipases, or proteases. Acid and neutral proteases have little or no activity, but some alkaline proteases are able to form appreciable numbers of lactic acid from PLA [19]. The hydrolytic activity of enzymes depends on many factors, including pH and temperature. A successful Proteinase K-catalyzed graphene-PLA digestion was recently reported by Manzanares-Palenzuela et al. [20] opening a new environmentally friendly and reproducible alternative for the surface activation procedure of PLA-based 3D printed electrodes. Solvent-free activation by laser ablation was recently proposed by Glowacki et al. [21].

This work aims to present and understand the electrochemically active surface area development of the CB-PLA through the controlled electrolysis, with the goal to increase the electrode's utility as an electrochemical sensor platform.

Various polarization protocols were studied to evaluate the optimum conditions for CB-PLA activation. First, the anodic and cathodic polarization depth were evaluated, for the PLA electrolysis carried out in acidic and alkaline media, discussing the resultant modification in surface chemistry and the activation mechanism. A separate study was conducted to elaborate the CB-PLA activation protocol by proteinase K enzymatic interaction. Finally, the efficiency of the synergistic interaction of the enzymatic hydrolysis and the electrolysis were reported. To the authors best knowledge, this is the first attempt to explore activation possibility through both of these processes, carried out simultaneously.

2. Experimental

**3D printout electrodes**: Flat electrodes with dimensions 10x10x2 mm were 3D-printed from conductive, commercially available PLA, Proto-Pasta, on an Ender 3 Pro 3D printer (Ender, China). The printing temperature was 200 °C. According to the thermogravimetric analysis (presented in the **Supplementary Information file, Section S1**), the CB-PLA filament contains 26.4 wt.% carbon filler, which results in 30 Ωcm electric resistivity. The filament, as well as the electrodes, were stored under atmospheric conditions.

**Activation protocols:** The activation procedure was started immediately after the printing process. Chemical, electrochemical and enzymatic activation protocols were investigated. The electrochemical activation was performed in cyclic voltammetry regime, applying the polarization at a constant rate of 50 mV/s in the below-defined polarization range. Each electrode was subjected to a total of 10 polarization cycles. The chemical and enzymatic activation treatments were carried out simply by soaking the electrode in the target electrolyte. Solvent treatments were performed by soaking the 3D-printed electrodes in one of the studied electrolytes for 24 hours. For determination of the effect of the enzyme activity on the CB-PLA, two tests were performed. In the first, each 3D-printed specimen was immersed in a 0.6 mg /mL proteinase K solution prepared in Tris-HCl buffer (Tris 100 mM, $CaCl_2$ 1 mM, pH 8.0) and incubated at 37°C for 24h, 48h, 72h and 96 h. In the second, each 3D-printed specimen was immersed in a proteinase K solutions with 0.2 mg/mL, 0.4 mg/mL, 0.6 mg/mL and 0.8 mg/mL prepared in Tris-HCl buffer (Tris 100 mM, $CaCl_2$ 1 mM, pH 8.0) and incubated at 37°C for 72 h. The enzyme/buffer was replaced every 24 h to restore enzymatic activity, avoiding the pH value decrease. After CB-PLA electrochemical activation was finished, the samples were rinsed thoroughly with distilled water at 4 °C to stop further degradation and then dried until constant mass. The mass loss (%) was calculated according to $(m_0 - m_t)/m_0$, where $m_0$ and $m_t$ represent the dry weights of the specimens before and after degradation, respectively.

**Electrochemical studies**: All the electrochemical measurements, including electrochemical activation and electrochemically active surface area (EASA) determination of the electrode, were carried out in a three-electrode setup with the 3D-printed working electrode, Ag|AgCl as the reference electrode and platinum wire as the counter electrode. The electrochemical measurements were done using a Reference 600+ potentiostat/galvanostat (Gamry Instruments, USA) and controlled by Gamry Framework software.

Determination of the surface activation efficiency was based on charge transfer kinetics evaluation in electrolyte composed of 2.5 mM $[Fe(CN)_6]^{4-/3-}$ as the electroactive redox species dissolved in 0.5 M $Na_2SO_4$. Two techniques were used, namely electrochemical impedance spectroscopy (EIS) and cyclic voltammetry (CV). The EIS experiment was carried out at open circuit potential (OCP) conditions (0.25 ± 0.04 V vs Ag|AgCl), with voltage perturbation amplitude of 10 mV, and in the frequency range of 100 000 to 0.1 Hz, 10 points per frequency decade. The CV experiments were done at a scan rate of 50 mV/s. The $[Fe(CN)_6]^{4-/3-}$ species were used as the most commonly referred redox-active probe. Moreover, the inner sphere electron transfer (ISET) mechanism of the ferrocyanides is known to be largely dependent on the electrode surface chemistry [22,23]. Thus, we found this redox probe to be particularly valuable to differentiate the efficiency of surface activation protocols.

**Physicochemical studies**: The X-Ray photoelectron spectroscopy (XPS) studies were carried out with Escalab 250Xi multispectroscope (Thermo Fisher Scientific, USA), equipped with monochromatic AlKα source (spot diameter 650 μm). The high-resolution spectra of CB-PLA electrodes after various surface activation routes were collected in the core-level binding energy range of *C 1s* and *O 1s* peaks. The pass energy was 15 eV. Charge compensation was controlled through a low-energy electron and low-energy $Ar^+$ flow. Next, the scanning electron microscopy (SEM) examination of surface activated samples was carried out using S-3400N VP-SEM microscope (Hitachi, Japan), using secondary electron mode and 20 kV accelerating voltage. Prior to the examination, the CB-PLA electrodes were sputtered with 10 nm gold layer.

## 3. Results and discussion

### 3.1. The CB-PLA electro-activation by PLA electrolysis

The efficiency of the electrochemical activation of each studied CB-PLA electrode, illustrated through the CV curves, is shown in **Fig. 1**. The CB-PLA electrodes were subjected to potentiodynamic polarization treatment under various applied anodic ($\eta_A$) or cathodic ($\eta_C$) overpotentials, aiming to evaluate the effect of the electrolysis on surface electro-activation efficiency.

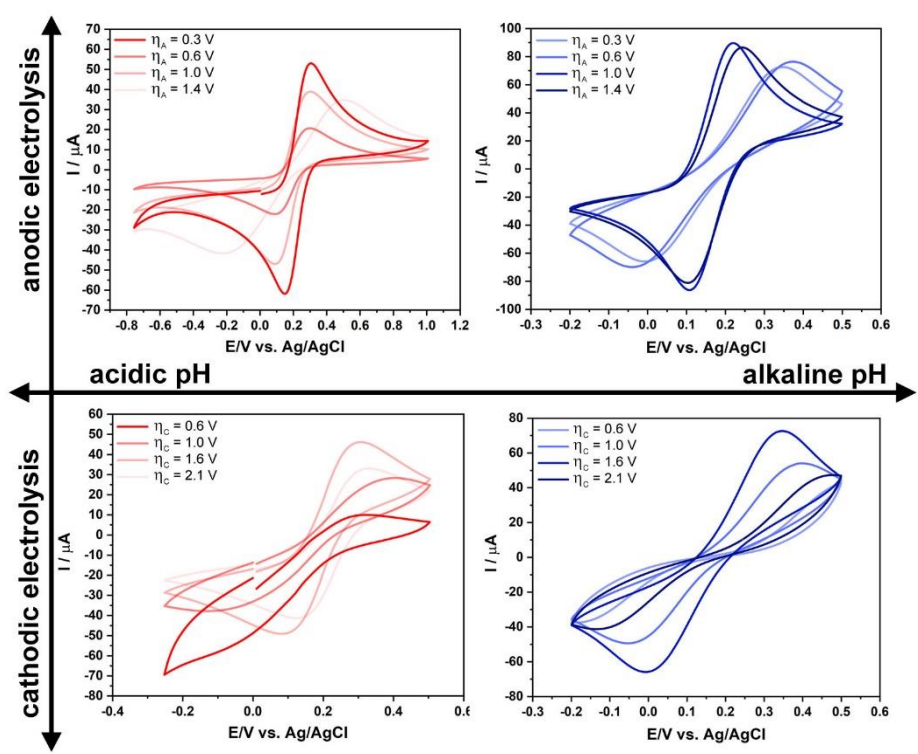

**Figure 1** – a) CV curves reflecting charge transfer kinetics through the electro-activated CB-PLA electrode depending on the direction and depth of electrochemical polarization, carried out in the electrolyte with acidic (1M HCl) and alkaline (1M NaOH) pH. Electrolyte: 0.5 M $Na_2SO_4$ + 2.5 mM $[Fe(CN)_6]^{4-/3-}$. Polarization scan rate 50 mV/s.

The results displayed in **Fig. 1** confirm that the electrochemical treatment significantly enhances the charge transfer kinetics at the electrode interface, proving their successful activation, both acidic and alkaline environments. Moreover, the charge transfer is strongly affected by the polarization direction and the polarization depth. When comparing the efficiency of the ferrocyanide/ferricyanide ions oxidation/reduction process, it should be concluded that exceptionally high efficiency is obtained by applying anodic overpotential at the electrode surface.

Ferrocyanides oxidation proceeds more efficiently at the CB-PLA electrode activated in the alkaline media than in acidic media. The reversibility of the studied process improves significantly in these conditions, testified by $\Delta E_P$ = 112 mV and $i_A:i_C$ = 0.96 ($\eta_A$ = 1.0 V, scan rate 50 mV/s). Given the CB-PLA activation is carried out at sufficiently high anodic polarization, the electrolysis leads to higher values of $k^0$ and more effective redox kinetics. This effect is observable in particular at $\eta_A$ exceeding 0.6 V, when the observable value of $k^0$ increases by order of magnitude, peaking at 0.003 cm/s, the exact values to be found in **Table 1**. The considerable improvement in electron charge transfer is probably due to removing the polylactide matrix and exposing conductive CB fillers at the electrode surface. The efficiency of the activation process is more evident in alkaline media, where the oxidation peak current reaches 90 µA after CB-PLA activation at the highest studied $\eta_A$. However, even when the activation was carried out at much smaller potentials, not exceeding +0.5 V vs. Ag|AgCl, the redox process kinetics was considerable, with $i_A$ = 66 µA. The standard heterogeneous rate constant $k^0$ was estimated using the Nicholson approach, with eq. (1) [24,25]:

$$k^0 = \psi \left(\pi D_0 \frac{nFv}{RT}\right)^{1/2} \qquad (1)$$

where $\psi$ is the kinetic parameter estimated based on peak potential separation $\Delta E_P$, $D_0$ is the diffusion coefficient (6.67*10$^{-6}$ cm$^2$/s), n is the number of electrons transferred, $v$ is the applied scan rate, R is the gas constant, F is the Faradaic constant and T is the temperature.

**Table 1** – The CV characteristics of the $[Fe(CN)_6]^{4-/3-}$ redox at CB-PLA electrodes, obtained after various polarization treatments in 1M HCl or 1M NaOH electrolyte, measured at polarization scan rate 50 mV/s.

|  | **1M HCl** | | | | **1M NaOH** | | | |
|---|---|---|---|---|---|---|---|---|
| **$\eta_A$ / V** | **0.3** | **0.6** | **1.0** | **1.4** | **0.3** | **0.6** | **1.0** | **1.4** |
| $\Delta E_P$ / mV | 443 | 282 | 192 | 151 | 356 | 413 | 112 | 140 |
| $i_A$ / µA | 37.8 | 42.6 | 23.2 | 57.5 | 65.9 | 69.9 | 86.3 | 81.2 |
| $i_A:i_C$ / - | 1.01 | 1.00 | 0.90 | 0.95 | 0.91 | 0.91 | 0.96 | 0.94 |
| $k^0$ / cm/s | 3.2e-4 | 5.1e-3 | 9.6e-4 | 1.4e-3 | 3.8e-4 | 3.2e-4 | 2.7e-3 | 1.6e-3 |
| **$\eta_C$ / V** | **0.6** | **1.1** | **1.6** | **2.1** | **0.6** | **1.1** | **1.6** | **2.1** |
| $\Delta E_P$ / mV | -- | 390 | 159 | 180 | 687 | 447 | 356 | 615 |
| $i_A$ / µA | 12.2 | 27.3 | 46.2 | 33.7 | 37.5 | 49.4 | 65.9 | 41.3 |
| $i_A:i_C$ / - | -- | 0.74 | 0.90 | 0.80 | 0.88 | 0.91 | 0.91 | 0.87 |
| $k^0$ / cm/s | -- | 3.8e-4 | 1.3e-3 | 9.6e-4 | 1.9e-4 | 3.2e-4 | 3.8e-4 | 1.9e-4 |

The highest value of the heterogeneous rate constant after activation in acidic media was two times lower and achieved after activation at the highest anodic overpotentials applied during the activation process, with oxidation peak current $i_A$ reaching 57.5 µA (at 50 mV/s scan rate). Nevertheless, the peak separation potential at the most efficient studied $\eta_A$ was $\Delta E_P$ = 151 mV, which, together with a slightly unbalanced anodic-to-cathodic current ratio $i_A:i_C$ = 0.89, hints at the irreversibility of the electrochemical process. Notably, the activation performed in both studied electrolytes may lead to obtaining the quasi-reversibility regime of the redox process [26].

A key observation should also be made based on the analyses of the above-presented results. The CB-PLA electrodes should be handled with great care since the polarization regime of commonly used CV procedures when using the most popular redox probes often exceed the polarization limits sufficient for CB-PLA surface modification throughout the measurement. On the other hand, most research groups dealing with 3D printed electrodes study the electron charge transfer using potassium hexacyanoferrate(II) as the redox probe. The impedimetric measurements may be found a more suitable approach to investigate the electrochemical response of the 3D-printed PLA-based electrodes, which is due to the mV polarization range of the perturbation amplitude.

The polarization curves registered during the surface activation treatment are presented in the **Supplementary Information file, Section S2.** The sole action of the studied electrolyte, in the absence of the electrode polarization, was shown in the **Supplementary Information file, Section S3**. The obtained results indicate that surface activation by PLA hydrolysis in absence of the polarization component is also the most effective when hydrolysis in carried out in the alkaline environment, providing a well-developed EASA and electron transfer rate. On the other hand, with very slow electron transfer kinetics of only partially-surface activated electrodes in acidic and neutral pH environments, the obtained results were poor, without defined redox peaks. The increased PLA susceptibility to saponification in alkaline solution should be explained by more effective PLA hydrolysis in these conditions and thus better exposure of the conductive CB nanoparticles at the electrode surface. Most notably, the surface activation in 1M NaOH under cyclic polarization conditions was found a time-efficient and environmentally-friendly alternative to etching in aprotic solvents such as DMF.

While the anodic activation treatment leads to the pronounced CB-PLA EASA development, performing the activation process at cathodic polarization to a certain extent increases the faradaic charge transfer. Yet, the activation efficiency is dramatically lower, and the resultant heterogeneous rate constant barely competes with the $k^0$ value obtained after activation at the lower studied anodic polarization. These results prove the smaller utility of cathodic polarization treatment as a part of the 3D-printed electrodes activation protocol [17,27]. Overall, the optimum electrochemical activation polarization range was determined to be from -1.4 V to +1.2 V vs. Ag/AgCl, as the smallest peak-to-peak separation and the highest redox probe faradaic current were achieved.

A similar observation regarding the CB-PLA EASA development efficiency was drawn based on the electrochemical impedance spectroscopy measurements of the studied electrodes. These results were carried out in a 1M NaOH solution, found highly effective for the activation process. The Nyquist plots are depicted in **Fig. 2**.

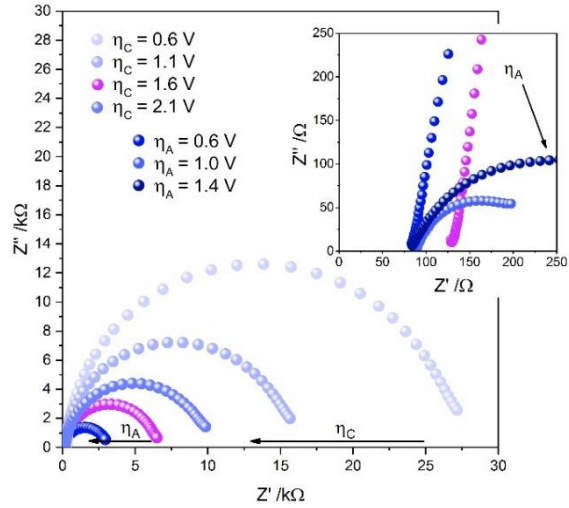

**Figure 2** – EIS spectra for CB-PLA electrodes electrochemically activated under different potentiodynamic polarization conditions, in 1M NaOH. Studies carried out in 0.5 M $Na_2SO_4$ + 2.5 mM $[Fe(CN)_6]^{4-/3-}$.

The impedance data were analyzed using an R(QR) electric equivalent circuit (EEC) built of a series resistance $R_S$ (electrolytic resistance) and a parallel connection of a charge transfer resistance ($R_{CT}$) and constant phase element (CPE) representing the electric double-layer capacitance of a heterogeneous electrode. The impedance of the CPE may be defined with eq. (2) [28].

$$Z_{CPE} = \frac{1}{Q(j\omega)^\alpha} \quad (2)$$

where Q is the quasi-capacitance, j is the imaginary number, and ω is the angular frequency. For α = 1, the CPE represents the ideal capacitor, and lower α values are introduced by frequency dispersion of capacitance due to heterogeneous charge transfer at the electrode/electrolyte interface [28,29]. The electric parameters from the fitting procedure with the selected EEC are given in **Table 2**.

**Table 2** – The electric parameters obtained for CB-PLA electro-activation in 1M NaOH with each studied electrochemical polarization condition.

|  | η / V | $R_{CT}$ / kΩ | Q / μSs$^n$ | α / - |
|---|---|---|---|---|
| $\eta_C$ | 0.6 | 27.41 | 2.47 | 0.93 |
|  | 1.1 | 15.85 | 4.08 | 0.94 |
|  | 1.6 | 6.41 | 5.57 | 0.95 |
|  | 2.1 | 9.92 | 4.97 | 0.93 |
| $\eta_A$ | 0.3 | 6.41 | 5.57 | 0.95 |
|  | 0.6 | 3.05 | 5.81 | 0.93 |
|  | 1.0 | 0.16 | 43.92 | 0.80 |
|  | 1.4 | 0.34 | 46.45 | 0.70 |

Altering of the cathodic polarization potentials during the electrochemical activation of CB-PLA electrodes influences the charge transfer resistance. Compared with the previously presented untreated CB-PLA electrode ($R_{CT}$ = 326 kΩ), the $R_{CT}$ parameter drops by one order of magnitude yet does not fall below 6 kΩ (98% efficiency) regardless of the applied cathodic polarization depth limit.

On the other hand, the deeper the anodic polarization range, the lower the $R_{CT}$ value. In particular, for $\eta_A > 0.6$ V, the $R_{CT}$ drops by another order of magnitude and reaches 160.3 Ω at $\eta_A = 1.0$ V, which exceeds 99.9% activation efficiency. Notably, the EIS-measured $R_{CT}$ characteristic of the surface-activated CB-PLA electrodes corroborates the CV-measured $i_A$ and $k^0$. An interesting feature was observed when analyzing the CPE-exponent α changes, suggesting that the electric heterogeneity of each cathodically-activated CB-PLA electrode surface is similar and only decreases if the activation procedure is carried out with deep anodic polarization limits ($\eta_A > 0.6$ V). There are few possible explanations of such characteristics. Most likely, highly efficient uncovering of the conductive CB particles leads to the appearance of the surface distribution of the time-constant dispersion, falling in the mechanism of heavily overlapping diffusion layers at spatially heterogeneous electrodes [30,31]. The second plausible cause is introducing the normal distribution of the time constants through the appearance of the porous electrode surface.

The high-resolution XPS spectra were recorded in the *C 1s* (**Fig. 3a**) and *O 1s* (**Fig. 3b**) core-level binding energy range to determine the CB-PLA surface chemistry resulting from various electro-activation protocols and PLA etching. The untreated, 3D printed CB-PLA electrode was also studied for comparison purposes.

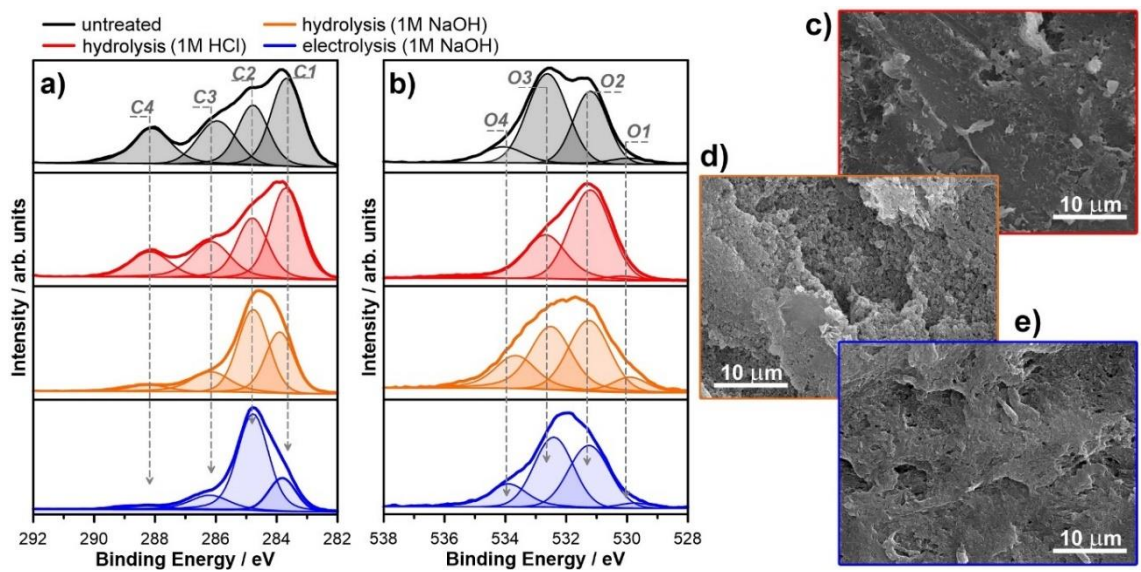

**Figure 3** – a,b) XPS spectra registered in a) C 1s and b) O 1s binding energy range for the untreated CB-PLA electrode and after its electro-activation using various protocols, c-e) SEM micrographs of CB-PLA electrode surface after hydrolysis in c) 1M HCl and d) 1M NaOH, and e) electrolysis in 1M NaOH (-1.4 to +1.2 V vs. Ag|AgCl polarization range).

The polylactide chemistry includes three different carbon chemical states, namely C-C, C-O, and C=O. These three types of chemical bonds are commonly identified in the *C 1s* spectra as signals peaking at a binding energy of 284.6 eV (C2), 286.0 (C3), and 288.1 eV (C4), respectively [21,32,33]. Furthermore, their expected ratio for PLA matrix should be 1:1:1. A similar ratio of C2:C3:C4 components was found for the untreated sample. The aliphatic C-C bonds (C2) dominated due to the simultaneous detection of surface adventitious carbon contaminants from atmospheric exposure [34]. The most prominent component (C1) recorded in the untreated PLA *C 1s* spectra lies in the energy range characteristic to the sp$^2$-carbon in CB [21,35,36], revealing that approx. 27 at. % signal originates from the composite filler, corroborating the thermogravimetric analysis. The data are presented in **Table 3**. The oxygen chemistry analyzed for the untreated sample may also be divided to four different types of interactions. The dominant two, O2 and O3, are characteristic of C=O and C-O bonds, respectively, originating primarily from the polylactide matrix. Two minor oxygen components were also identified, first the O1 at 529.9 eV, a value typical for metal oxides, impurities to commercially available PLA filaments originating from thermal stabilizers and plasticizers [37]. Finally, oxygen in chemisorbed water molecules may result in the signal appearing at approx. 534 eV, identified as O4 [38,39].

**Table 3** – The CB-PLA surface chemistry (in at.%) in terms of various chemical states of carbon and oxygen, based on the XPS analysis with the deconvolution model.

|  |  | *C 1s* | | | | *O 1s* | | | |
|---|---|---|---|---|---|---|---|---|---|
|  |  | C1 | C2 | C3 | C4 | O1 | O2 | O3 | O4 |
| **BE/ eV** |  | 283.7 | 284.6 | 286.0 | 288.1 | 529.9 | 531.2 | 532.7 | 534.1 |
| **Untreated** |  | 27.0 | 18.6 | 15.9 | 14.6 | 1.0 | 8.4 | 12.2 | 2.4 |
| **Hydrolysis** | HCl | 26.5 | 18.0 | 13.9 | 9.1 | 0.9 | 19.8 | 11.3 | 0.5 |
|  | NaOH | 29.2 | 41.4 | 14.4 | 4.2 | 0.9 | 4.0 | 3.9 | 2.1 |
| **Electrolysis** | NaOH | 16.0 | 55.6 | 11.1 | 2.7 | 0.5 | 5.7 | 6.2 | 2.3 |

Both carbon and oxygen chemistry are significantly affected by the applied CB-PLA surface activation protocols, yet the mechanism of the interaction based on the obtained XPS results is complex and possibly different in each case. The hydrolysis process, carried out in an aqueous 1M hydrochloric acid solution, appears to lead to a significant increase in the share of oxygen content. 24 at. % for the untreated CB-PLA electrode up to 32.3 at. %. On the other hand, due to the PLA treatment in the alkaline environment, the amount of surface oxygen atoms is twice reduced. This observation is visible with and without activation by electrolysis. The above-described change is primarily recognized as the increase of C-O and C=O interactions, while the share of O1 and O4 components is negligibly affected by the chemical oxidation process and does not exceed 1.4 and 2.4 at.%, respectively. This effect will be discussed further on.

The exposure of the electrically conductive carbon black filler at the CB-PLA surface is evident based on the increase of charge transfer kinetics through the activated electrode. Yet, this observation is not assisted by the rise in *C 1s* carbon black nanofiller C1 component share. This is an important yet unexpected observation, confirmed by each studied activation procedure, either through PLA hydrolysis or electrolysis. The possible explanation of this mechanism may originate from partial oxidation of the sp$^2$-carbon filler material. The excessive oxidation of the electrode surface during the electrolysis process was presumably the cause of more irreversible charge transfer observed after anodic electrochemical activation of 3D printed CB-PLA electrodes by Vaněčková et al. [40]. According to this hypothesis, the amount of CB exposed at the electrode surface increases, yet its oxidation leads to a positive binding energy shift towards energies characteristic for C2 or even C3 components, superimposing with PLA-based features. A possible confirmation of the above-presented explanation may be found in our recent XPS examination of the surface PLA treatment by laser ablation in air and helium atmosphere, leading to CB-PLA electrochemical activation. Laser ablation in the air atmosphere is characterized by a nearly 25% decrease of C1 with a similar increase in C2 peak share. On the other hand, the noble gas matrix does not lead to CB oxidation, resulting in an over 20% C1 increase and only a 7% C2 increase compared with the results of the untreated CB-PLA electrode [21].

The optional explanation of the above-defined phenomenon is the rapid surface area development of the CB-PLA electrode due to PLA hydrolysis process, leading to breaking the ester bond and an increase in the amount of aliphatic C-C bonds at the surface, confirmed with a rapid decrease of C=O bonds (C4, O2 peaks), in particular in an alkaline environment. The development of the surface area will also consequence in the increased amount of surface-adsorbed adventitious carbons. Surface area development is visible on SEM micrographs, in **Figs. 3c-e**. The possible mechanism of the PLA hydrolysis is presented in **Fig. 4**. The decrease in C1 share may also be attributed to the detaching of loosely linked conductive carbon nanoparticles by PLA hydrolysis or the binding energy shift resulting from the disappearance of C-C bonds between the CB and the polymer [41]. Notably, the above-proposed hypotheses are not excluding each other, and both may co-occur, however, their verification requires independent studies.

The hydrolysis of the PLLA (poly(L-lactic acid) ester bonds occurs in acidic and alkaline environment. It is well known that the hydrolytic degradation of ester bonds is slow in aqueous conditions, while it is much faster in alkaline environments than in acidic ones [42]. The efficiency of PLLA hydrolysis is very slow in neutral pH and moderate acidic and alkaline solutions. Therefore, the activation of the CB-PLA surface was performed in strongly acidic pH 0 and alkaline conditions at pH 14. It is worth noting that in pH 14, the hydrolytic degradation is much faster than in pH 0. The mechanism of PLA hydrolysis previously proposed by Lucas and co-workers [43] involves two paths of degradations by intrachain or end chain of PLA regardless of the environment.

The mechanism of PLA hydrolysis performed in acidic solution involves the protonation of hydroxyl groups end groups, which lead to the formation of intramolecular hydrogen bonds and contribute to the hydrolysis reaction and the formation of free lactic acid, causing the decrease the length of PLA. On the other hand, the acidic conditions lead to the random protonation of the carbonyl group at the oxygen of PLA leads to the intramolecular hydrolysis of ester bonds in the PLA chain. This reaction cause the degradation of polymer and the formation of PLA chains consisting of lower molecular weights (Fig. 4a). The hydrolytic degradation of PLA also occurs in a strongly alkaline solution, where it involves intramolecular degradation and a reaction at the polymer end groups. The possible mechanism of end chain PLA degradation, including intramolecular transesterification, is suggested by Jong [44] (Fig. 4b). In alkaline conditions, nucleophilic attack of the hydroxyl end group on the carbonyl group causes the formation of a six-membered ring as an intermediate. The newly formed free lactide hydrolyzes into two species of lactic acid. However, the intramolecular chain is randomly hydrolyzed as hydroxide ion attacks the carbonyl groups of esters leading to hydrolysis. In this reaction path, two new molecules are formed.

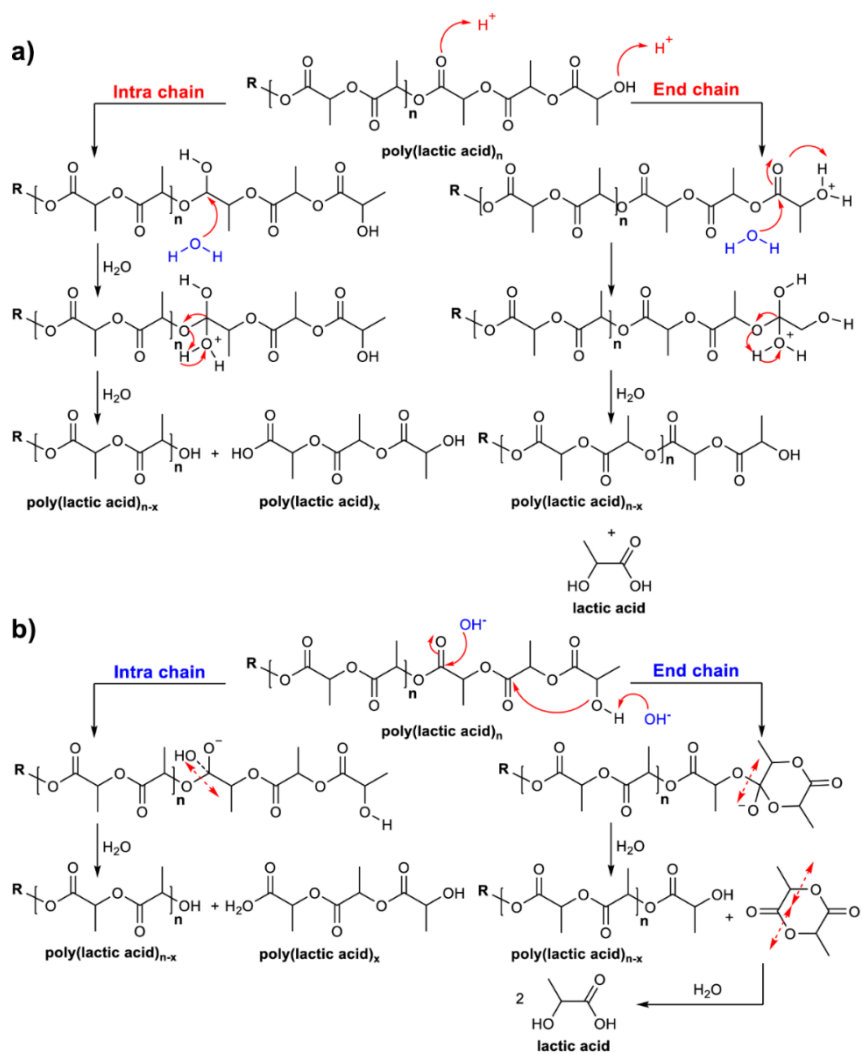

**Figure 4** – The mechanism PLA hydrolysis in a) acidic, b) alkaline environment.

The results of the XPS studies confirm that during the electrochemical degradation in acidic solution, the significant increase in the share of oxygen content, from approx. 24 at. % for the untreated CB-PLA electrode up to 32.3 at.% is observed, which can be a consequence of both the intramolecular hydrolysis and hydrolysis occurring at the end groups. Furthermore, the hydrolysis performed in an alkaline solution cause that the share of surface oxygen atoms is two times lower, which suggests that intramolecular hydrolysis is preferred.

The alteration of the electrode polarization causes the alkalization of the electrode surface in the anode area and acidification in the cathode area. Additionally, the solution pH has a significant influence on the alkalization or acidification at the electrode surface. The total effect, dependent on the implementation of electrochemical process conditions, can lead to a local pH change towards effective acidic or alkaline hydrolysis of PLA. Therefore, the overall efficiency depends on the sum of both overlapping processes. Thus, the applied polarization significantly influences the PLA hydrolysis, as shown in **Figure 1**.

### 3.2. The CB-PLA electro-activation by synergistic electrolysis and enzymatic hydrolysis interaction

Proteinase K is an enzyme to catalyze the hydrolysis of PLLA (poly(L-lactic acid) since the structure of the PLLA monomer is similar to alanine [45]. The hydrolytic activity of enzymes depends on many factors, including pH and temperature. The critical parameter is also the chirality of the lactide unit, which exists in three diastereoisomeric forms: L-lactide (PLLA), D-lactide (PDLA) and meso-lactide [46]. Both PLLA and PDLA are enzymatically hydrolyzed by two different classes of enzymes: proteases and lipases. One of the best known efficiently hydrolyzing PLA alkaline proteases is proteinase K. A successful Proteinase K-catalyzed graphene-PLA digestion was recently reported by Manzanares-Palenzuela et al. [20], opening a new environmentally friendly and reproducible alternative for the surface activation procedure of PLA-based 3D printed electrodes. These precious studies point the possibility of CB-PLA activation through enzymatic hydrolysis, however, the prolonged treatment duration hinders the possible laboratory applications of this approach.

Due to the reported long time required for the enzymatic hydrolysis, we have decided to verify the possibility of achieving the synergistic interaction between enzymatic and electrochemical PLA surface activation for a more time-efficient protocol. In order to do so, first, we have studied the enzymatic-driven surface activation of the carbon black-PLA electrodes, achieving the tailorable electrode performance, dependable on the Proteinase K digestion conditions. The results of the enzymatic hydrolysis on the CB-PLA electrode are summarized in **Fig. 5**.

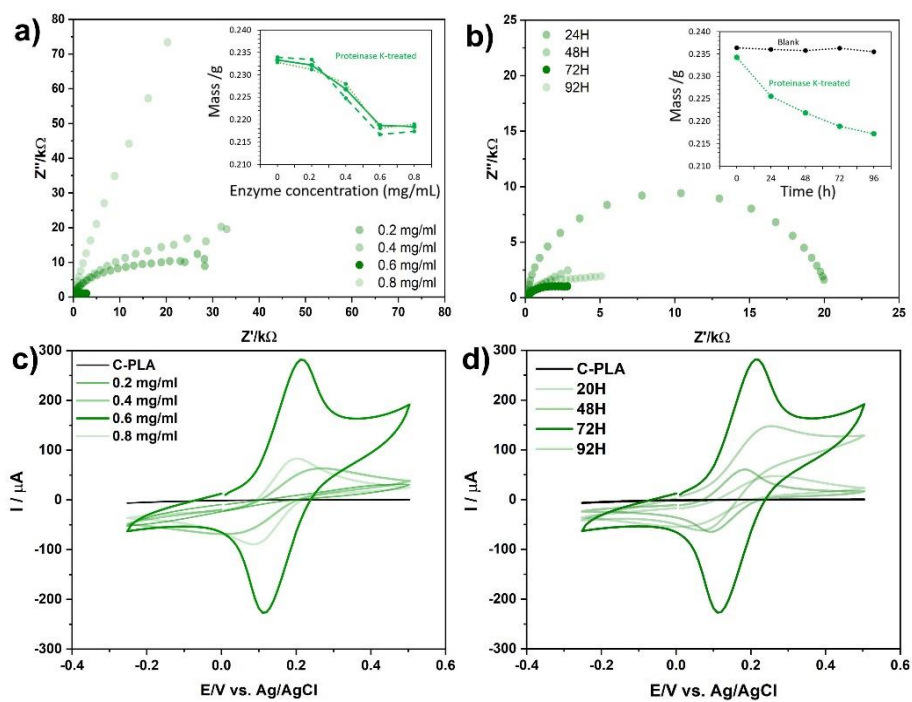

**Figure 5** – a,b) CV curves and c,d) EIS spectra reflecting charge transfer kinetics alteration, depending on various studied proteinase-K digestion conditions, i.e. a,c) enzyme concentration and b,d) enzymatic hydrolysis time. Mass variation during CB-PLA electro-activation in the inset: a) 72 h activation, b) 0.6 mg/mL proteinase-K concentration.

Increasing the proteinase K concentration from 0.2 mg/mL to 0.6 mg/mL improved the electrode surface activation effect. The peak separation of the ferrocyanide redox probe was reported as 91 mV at 0.6 mg/mL proteinase K concentration, allowing to estimate the heterogeneous rate constant $k^0$ as 0.05 cm/s, almost twice the value obtained previously, in the case of the most effective electrochemical activation protocol. Likewise, the faradaic oxidation currents $i_A$ significantly exceeded CB-PLA values after electrochemical activation protocol (86.3 μA), reaching 222 μA (at 50 mV/s scan rate) for proteinase K concentration of 0.6 mg/mL. These parameters are reported in **Table 4**. However, at higher concentrations (0.8 mg/mL), the reaction will no longer speed up since the amount of substrate was available to all the enzyme active sites at the electrode surface, reaching the rate-limiting factor. Thus, the optimal concentration of the enzyme in the tested system turned out to be 0.6 mg/mL. Increasing enzyme concentration will speed up the reaction. However, once all substrates are bound, the reaction will no longer speed up since there will be nothing enzymes may bind to [47].

**Table 4** – The CV characteristics of the $[Fe(CN)_6]^{4-/3-}$ redox at CB-PLA electrodes, obtained at various proteinase K digestion conditions, measured at polarization scan rate 50 mV/s.

| | concentration / mg/mL | | | | activation time / h | | | |
|---|---|---|---|---|---|---|---|---|
| $\eta_A$ / V | 0.2 | 0.4 | 0.6 | 0.8 | 20 | 48 | 72 | 92 |
| $\Delta E_P$ / mV | -- | 201 | 91 | 111 | 221 | 250 | 91 | 141 |
| $i_A$ / μA | -- | 54.5 | 222.4 | 87.0 | 44.4 | 60.2 | 222.4 | 80.9 |
| $i_A:i_C$ / - | -- | 1.04 | 0.96 | 1.04 | 1.06 | 1.01 | 0.96 | 1.00 |
| $k^0$ / cm/s | -- | 7.7e-4 | 4.8e-3 | 2.9e-3 | 5.8e-4 | 5.4e-4 | 4.8e-3 | 1.6e-3 |

The CV studies find confirmation in the shape of the impedimetric spectra. At the lowest proteinase K concentration, the $R_{CT}$ exceeded 300 kΩ and dropped significantly, with the enzyme concentration increase reaching 20.3 kΩ at 0.6 mg/mL. However, the shape of the impedance spectra after enzymatic activation was more complex than previously reported electrochemical treatment, revealing a second time-constant most likely originating from the surface-adsorbed proteinase K layer. These results corroborate the mass loss measurement presented in the inset of **Fig. 5a**, where the maximum mass loss of about 170 mg after 72 h exposure was consistently reported after digestion in 0.6 mg/mL proteinase K concentration, with the plateau observable as a result of PLA digestion at higher enzyme concentrations.

The CB-PLA mass-loss studies after enzymatic action in the inset of **Fig. 5d** reveal the PLA digestion rate with the exposure duration. Notably, the highest mass loss was observed during the first day of activation and slowly reaching the plateau at higher exposure times. Nevertheless, even after 72 h of exposure, further digestion process is still observable. This result provides another proof of different 3D printed PLA-based electrode surface activation by proteinase K depending on the conductive carbon filler since in the original studies, longer digestion times (72 h) disintegrated the 3D-printed electrodes [48]. In the case of CB-PLA, no such behavior was observed. On the contrary, the highest surface activation efficiency measured with heterogeneous rate constant $k^0$ and faradaic oxidation current $i_A$ values was incomparably smaller in the case of smaller activation durations.

The short, one-day-long activation procedure as reported by Manzanares-Palenzuela et al. [48] was found of little efficiency in the case of CB-PLA. Even at 72 h exposure, the reported proteinase K concentration 0.2 mg/mL was insufficient to reveal the faradaic currents originating from the redox probe, suggesting that the CB-PLA composite shows significantly lower activation efficiency than originally used graphene-PLA. The proteinase K-catalyzed degradation rate is determined by the type, concentration, shape, dimension, dispersion, and adhesion of fillers [49]. For example, fullerene and carbon nanotubes accelerated the enzymatic degradation due to the creation of large gaps between PLA phase and facilitation of the proteinase K diffusion into the material (inside and surface cleaving). Because of the lack of adhesiveness of the fillers mentioned above to the PLA phase, they should have been readily released from the film surface, forming a porous structure, increasing the surface area for the action of the enzyme (facile release). They also confirmed that conventional carbons have minimal effects on the enzymatic degradation rate [49]. Thus, carbon black probably forms small gaps in the PLA matrix and releases poorly from its surface.

The XPS analysis (**Fig. 6**) reveals the progressive digestion of the PLA by the proteinase K. One can notice that enzymatic hydrolysis duration leads to the decreased share of the CB NP's (C1 component), reaching 14 at.% at the end of a 72 h digestion process, a similar trend to the one presented for the most efficient alkaline electrolysis experiment. This effect is assisted by a significantly decreased share of the oxidized carbon bonds, C-O (C3, O3), C=O (C4, O2). Moreover, the exposition to proteinase K leads to a small share of C-N bonds at the electrode surface, labeled as C 1s C5 peak at 287.2 eV and N 1s N1 peak at 399.0 eV. These peaks were ascribed to $sp^2$ hybridized N bonded to three atoms, C-N(-C)-C or C-N(-H)-C [50,51]. The total share of these species based on the C5+N1 sum peaks at 3.5 at.% and is independent of the studied proteinase K digestion duration. It was confirmed that the enzymatic action does not lead to excessive surface oxidation as the combined share of C3, C4, C5, O2, O3 peaks does not exceed 21 at.% compared to 26 at.% for activation in alkaline media (either through electrolysis or hydrolysis) and over 54 at.% for hydrolysis in acidic media. Thus, it is most plausible that the rapid increase in C2 peak share after successful surface activation is due to the surface area development rather than conductive carbon oxidation. The complete XPS analysis is summarized in **Table 5**.

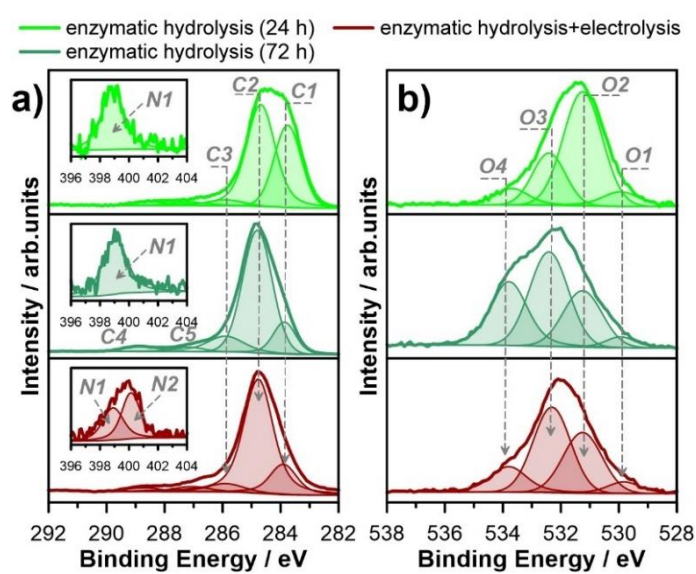

**Figure 6** – XPS spectra registered in a) C 1s and b) O 1s binding energy range for the CB-PLA electrode after activation in proteinase K for 24 and 72 h, and after combined action of electrolysis (-1.4 to +1.2 V vs Ag|AgCl polarization range) and enzymatic hydrolysis. The N 1s spectra in the inset.

**Table 5** – The CB-PLA surface chemistry (in at.%) after enzymatic hydrolysis in terms of various chemical states of carbon, oxygen and nitrogen, based on the XPS analysis with the deconvolution model.

|  | | *C 1s* | | | | *O 1s* | | | | *N 1s* | |
| --- | --- | --- | --- | --- | --- | --- | --- | --- | --- | --- | --- |
|  | | C1 | C2 | C3 | C5 | C4 | O1 | O2 | O3 | O4 | N1 | N2 |
| **BE/ eV** | | 283.7 | 284.6 | 286.0 | 287.2 | 288.5 | 529.9 | 531.2 | 532.7 | 533.8 | 399.0 | 400.1 |
| **Enzymatic hydolysis** | 24 h | 31.8 | 43.5 | 6.0 | 2.2 | 1.8 | 1.1 | 7.8 | 3.0 | 1.2 | 1.5 | -- |
|  | 72 h | 14.0 | 62.1 | 9.3 | 2.1 | 2.8 | 0.4 | 2.2 | 3.6 | 2.5 | 1.1 | -- |
| **Synergistic action** | | 14.1 | 57.9 | 5.4 | 3.2 | 1.9 | 0.9 | 4.8 | 6.5 | 2.2 | 1.4 | 1.7 |

The primary mechanism of proteinase K digesting of PLA polymer involves using a nucleophilic residue to cleave an ester bond. Serine proteases use an active serine to perform a nucleophilic attack on carbon of the ester group. Proteinase K degraded amorphous and homo crystalline regions of PLA films to produce oligo(lactic acid)s consisting mainly of linear ones and including small amounts of cyclic forms [52,53]. The mechanism of the enzymatic degradation depends on the enzyme's ability to recognize protein homologs, e.g. the lactic acid analogy in PLLA polymer to the L-alanine abundant in silk fibroins. Moreover, it is known that proteases can hydrolyze n-butyl or ethyl-D- and L-lactate but cannot PDLA, which is probably due to the problem with PDLA accommodation in the active site or making an acyl-enzyme intermediate. In PLLA the free electron pair present on the serine oxygen can interact with the carbon of the ester group and induce digestion. In the PDLA isomer, the methyl group creates a steric hindrance that blocks the substrate from accessing the enzyme's active site [54].

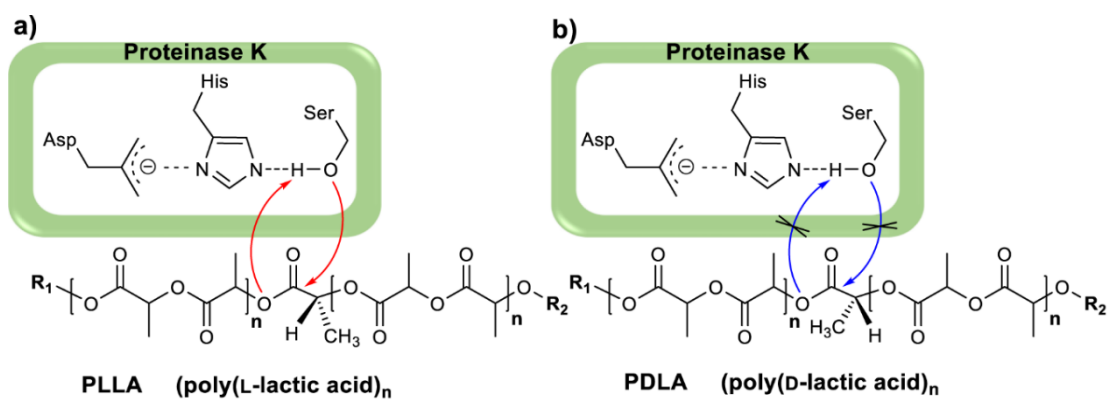

**Figure 7** – The mechanism of Proteinase K action in the presence of a) PLLA and b) PDLA.

The proposed mechanism was previously described by Kawai [36] and lead to the random cleavage of ester bonds in PLLA in a relatively short time (**Fig. 7a**), while this process is not observed in the presence of PDLA (**Fig. 7b**). Proteinase K is a serine protease with the classic catalytic triad of $Asp^{39}$-$His^{69}$-$Ser^{224}$ at its active site [55]. In the presented mechanism (Ser) is the primary nucleophile, while (His) plays a dual role both as a proton acceptor and donor at different steps in the reaction. The primary function of (Asp) is based on bringing the (His) residue in the correct orientation to facilitate nucleophilic attack by (Ser) [56]. The lack of hydrolysis observed in the PDLA results from the stereochemistry mismatch of the catalytic triad and PDLA (poly(D-lactic acid) or impossibility to make an acyl-enzyme intermediate.

Prolonged activation duration of CB-PLA electrodes under proteinase K digestion is unattractive when considering this activation protocol to be used effectively in laboratory studies. Therefore, we have verified the applicability of electrochemically-catalyzed enzymatic action to shorten the activation duration. Due to their chemical structure, enzymes are proteins. Like other proteins, they are endowed with a net charge, which is determined by the quantitative and qualitative composition of amino acids in the primary structure, spatial arrangement in higher-order structures, and the conditions of the environment in which it is located. The action of an electric field in an enzyme solution may cause the conformational change of the active site, the inhibition of the binding of substrate to protein, and the destabilization of the protein structure, which ultimately decreases the protein's activity [57].

The synergistic action of enzymatic hydrolysis and electrolysis on the CB-PLA etching was studied at elevated temperature (37 °C) and in Tris-HCl buffer (pH = 8), necessary for successful enzyme incubation. The results of this treatment are shown in **Fig. 8**. In addition, the effect of the electrochemical CB-PLA activation in Tris-HCl buffer at room temperature and at 37 °C was studied, and results presented in **Supplementary Information file, Section S4**. In the absence of proteinase K, the electrolysis process gave sub-optimal surface activation efficiency, with barely distinguishable redox processes and negligible effect of temperature increase.

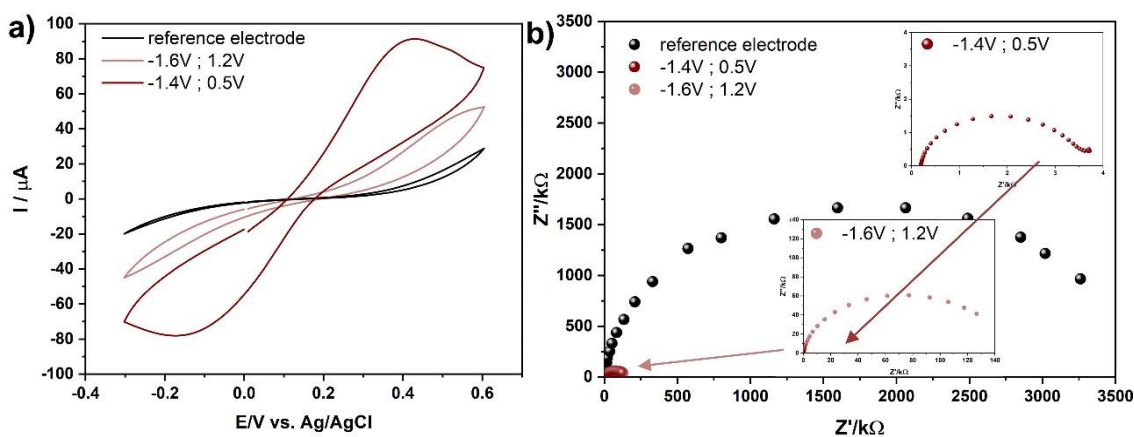

**Figure 8** – The results obtained for the CB-PLA electrode subjected to electrolysis in the presence of proteinase K: a) CV curves (scan rate 50 mV/s) and b) EIS spectra reflecting the charge transfer at CB-PLA electrode carried out at narrow and wide polarization range.

The enzymatic hydrolysis under the previously optimized polarization regime negligibly affects very poor CB-PLA electrode kinetics. Although the nonlinear Butler-Volmer characteristics were now observable in the CV studied polarization range, the redox probe process was still indistinguishable, assisted with a decrease in the charge transfer resistance to approx. 150 k$\Omega$. Oxidation and reduction peaks appear on the potentiodynamic polarization scan during the activation process (**Supplementary Information file, Fig. S5a**). The reduction peak at -1.2 V vs. Ag|AgCl may be explained by immobilizing the proteinase K on the CB-PLA surface [58]. In contrast, the surface modification is oxidized at +0.9 V. Since the process is irreversible, the corresponding faradaic currents develop with consecutive polarization scans.

A second experimental approach was carried out in a more narrow potentiodynamic polarization range with the anodic overpotentials not exceeding the value of the anodic polarization peak (CV scans upon $\eta_A$ = 0.3 V, seen in **Supplementary Information file, Fig. S5b**) to avoid removal of the proteinase K-functionalized layer. Performing the polarization in these conditions allowed us to observe prominent oxidation/reduction peaks from the ferrocyanide redox probe. The measured $\Delta E_P$ = 632 mV (scan rate 50 mV/s) hints at the irreversibility of the electrochemical process with the heterogeneous rate constant of approximately 4.0*10$^{-5}$ cm/s. Nevertheless, the irreversible electrode kinetics, the electrochemically active surface area (EASA) by carbon black is well-developed under proteinase K digestion and electrolysis co-occurrence. The direct proof of significant EASA development is the high value of the oxidation faradaic current $i_A$, exceeding 90 µA at the studied scan rate, similar to surface activation in 1M NaOH electrolyte. The charge transfer resistance measured in the optimized enzymatic-electrochemical activation conditions was not less than 3.5 kΩ. The conclusion should be drawn, attributing the enhancement of the ferrocyanide kinetics and CB-PLA EASA development to the surface modification during proteinase K electrografting [58].

The XPS analysis was carried out to study the surface chemistry of the sample after the combined enzymatic hydrolysis and electrolysis ($\eta_A$ = 0.3 V). The surface carbon chemistry reveals a high resemblance to the previously studied electrodes after enzymatic hydrolysis (72 h) and alkaline electrolysis due to the similar electrochemical surface area development level as testified by the electrochemical EASA examination. The essential differences observed are the decreased share of the C-O bonds (C3 drops by a factor of two, down to 5.4 at.%) and an increased amount of C-N species (C5+N1 increases by half, reaching 4.6 at.%). Moreover, another form of nitrogen appears, N2, peaking at 400.1 eV, attributed to hydrogen-bonded/protonated $NH_2/NH_3^+$ amine species [50]. The total nitrogen share grows twice. All the above observations hint at the electrodeposition by proteinase K.

The studies carried out on the combined enzymatic and electrolysis action towards the increase of CB-PLA electrochemical activation confirm that simultaneous activation of the CB-PLA electrodes via this modification route may be the new valuable direction of enhancing enzyme activity which requires a separate research path.

## 4. Final Remarks

The CB-PLA electrolysis by potentiodynamic polarization is the most efficient when carried out in an alkaline environment, delivering higher EASA and improved electron transfer rate. In particular, for anodic overpotentials reaching $\eta_A$ = 1.2 V the oxidation peak current $i_A$ of 86.3 µA (at 50 mV/s scan rate) and the peak-to-peak separation of $\Delta E_P$ = 112 mV were recorded. At the same time, the surface chemistry analysis by XPS revealed a significant decrease in oxygen content down to 12 % when compared with the untreated surface (32.3 at. %). However, the surface activation is less affected when the cathodic polarization range is modified, with the most efficient activation achieved at $\eta_C$ = 1.6 V. The solution pH significantly influences the alkalization or acidification of the electrode surface, which is further enhanced by changes in electrode polarization. The total effect, dependent on the implementation of electrochemical process conditions, can lead to a local pH change towards effective acidic or alkaline hydrolysis of PLA. The hydrolytic decomposition mechanism of PLA in strong alkaline conditions was attributed to the end chain degradation by intramolecular transesterification.

The enzymatic hydrolysis by proteinase K is less efficient for CB-PLA than in the case of previously reported graphene-PLA composite. The difference most likely originates from small gaps between CB and PLA phases, hindering proteinase K diffusion. The optimized digestion conditions were as follows: 72 h exposure at a 0.6 mg/mL proteinase K concentration, which resulted in very high ferrocyanide oxidation currents $i_A$ = 222 µA and higher reversibility ($\Delta E_P$ = 91 mV). Our results allow reporting enzymatic hydrolysis to be a significantly more effective protocol for CB-PLA electrochemical activation than electrolysis in either NaOH or HCl. However, prolonged treatment duration hinders the possible laboratory applications of this approach.

Finally, we have demonstrated a novel strategy of electro-activation of 3D-printed carbon black/poly(lactic acid) electrodes based on the simultaneous surface treatment by electrolysis and enzymatic hydrolysis with proteinase K. This allows for a unique synergistic interaction of electrolysis and enzymatic hydrolysis tailoring both electrochemically active surface area and electron transfer kinetics. This activation protocol has led to EASA development comparable to alkaline electrolysis activation ($i_A$ = 90 µA), and irreversible redox process. Furthermore, the XPS analysis reveals a high resemblance of the CB-PLA electrodes after enzymatic hydrolysis (72 h) and alkaline electrolysis, thanks to analogous electrochemical surface areas. The enhanced charge transfer kinetics was attributed to proteinase K electropolymerization under the cathodic currents.

The optimized values of the electrochemical parameters obtained for each studied activation protocol were summarized in **Table 6**.

**Table 6** - Optimized CV characteristics of the $[Fe(CN)_6]^{4-/3-}$ redox at CB-PLA electrodes after various applied activation protocols, measured at polarization scan rate 50 mV/s.

| Agent | 1M NaOH | 1M HCl | Proteinase K, 0.6 mg/mL | Proteinase K, 0.6 mg/mL |
|---|---|---|---|---|
| Polarization range (V vs Ag\|AgCl) | -1.4 ÷ +1.2 | -1.4 ÷ +1.2 | -- | -1.6 ÷ +0.5 |
| Modification | Electrolysis | Electrolysis | Enzymatic hydrolysis | Electrografting |
| $\Delta E_P$ / mV | 112 | 151 | 91 | 632 |
| $i_A$ / µA | 86.3 | 57.5 | 222.4 | 92 |
| $i_A$:$i_C$ / - | 0.96 | 0.95 | 0.96 | 1.17 |
| $k^0$ / cm/s | 2.7e-3 | 1.4e-3 | 4.8e-3 | 4.0e-5 |

**Acknowledgements:** This work was supported by The National Science Centre (Republic of Poland) under project SONATA BIS number 2020/38/E/ST8/00409 and by The National Centre of Research and Development (Republic of Poland) under project POLNOR number NOR/POLNOR/UPTURN/0060/2019-00.